\documentclass[prb,preprint, groupedaddress, showkeys, showpacs]{revtex4}

\usepackage{graphicx}
\begin{document}


\title{Transformation between dense and sparse spirals in symmetrical bistable media}


\author{Ya-feng He$^{1,2}$}
\author{Bao-quan Ai$^{1,4}$}
\author{Bambi Hu$^{1,3}$}


\affiliation{$^1$Department of Physics, Centre for Nonlinear
Studies, The Beijing-Hong Kong-Singapore Joint Centre for Nonlinear
and Complex Systems (Hong Kong), Hong Kong Baptist University,
Kowloon Tong, Hong Kong, China\\ $^2$College of Physics Science and
Technology, Hebei University, Baoding 071002, China\\$^3$Department
of Physics, University of Houston, Houston, TX 77204-5005, USA\\$^4$
Laboratory of Quantum Information Technology, ICMP and
 SPTE, South China Normal University, Guangzhou, China.}


\date{\today}
\begin{abstract}
\indent  Transformation between dense and sparse spirals is studied numerically based on a bistable FitzHugh-Nagumo model. It is found that the dense spiral can transform into two types of sparse spirals via a subcritical bifurcation: Positive Phase Sparse Spiral (PPSS) and Negative Phase Sparse Spiral (NPSS). The choice of the two types of sparse spirals after the transformation is affected remarkably by the boundary effect if a small domain size is applied. Moreover, the boundary effect gives rise to novel meandering of sparse spiral with only outward petals.
\end{abstract}



\maketitle
\section {Introduction}
\indent  Pattern formation has been of great interest in a variety of physical, chemical, and biological contexts. \cite{Cross, Koch, Epstein,  Petrov, He} Of particular interests are spiral waves because more and more evidence indicates that sudden cardiac death is related with the spiral waves in heart. \cite{Nattel} From the first observation in Belousov-Zhabotinsky reaction, spiral waves had been widely studied in excitable, bistable, and oscillation media. \cite{Berenstein, Zhang, Chen1, Mikhailov, Agladze, Li, Sheintuch, Bar1, Bar2} Most researches of spiral waves are focused on the excitable systems with different excitabilities. \cite{Grill, Karma, Margerit} Generally, spiral waves exhibit two types of appearances: dense spiral and sparse spiral. \cite{Krinsky, Xu} Dense spiral waves occur in media with normal excitability while sparse spiral waves appear in media with low excitability. The spiral waves always appear with the form of pulses because excitable media cannot support a single front. However, in bistable media a front can exist individually. The spiral wave in bistable media consists of a couple of Bloch fronts which propagate in opposite directions. An analytical relation about the velocity and the curvature of the Bloch fronts was obtained. \cite{Bar2} Numerical simulations shown that dense (sparse) spiral occurs in a symmetric (asymmetric) bistable system. \cite{Hagberg0, Hagberg1} In oscillation media the spiral waves are phase waves and exhibit dense spirals.

\indent In a bistable ferrocyanide-iodate-sulfite reaction, spirals, oscillating spot, and labyrinthine patterns have been observed. \cite{Li2, Szalai, Lee1, Lee2} The spirals occur in the Bloch region beyond the Nonequilibrium Ising-Bloch (NIB) bifurcation. It is sparse spiral and results from an axisymmetry breaking of a shrinking ring. The oscillating spots appear near but before the NIB bifurcation. The labyrinthine pattern originates from transverse instability of a chemical front in Ising region. The similar patterns were also observed by Kepper $et$ $al$ \cite{Szalai}. However, they obtained dense spiral. The observed dynamics of patterns can be explained successfully in terms of a NIB bifurcation in a generic FitzHugh-Nagumo model \cite{Hagberg0, Hagberg1}.

\indent Dense and sparse spiral waves have been investigated individually in those media. The behaviors of both dense and sparse spirals have been well understood. However, the transformation process between dense and sparse spirals is still unclear. In the present work we will focus on this transformation by studying numerically a symmetrical bistable FitzHugh-Nagumo model. The  spiral waves are obtained in Bloch region. It is found that the dense spirals can transit into two types of sparse spirals. We also observe novel meandering of sparse spiral with only outward petal which originates from boundary effects.

\section{BISTABLE MEDIA MODEL}

\indent This work is based on a modified FitzHugh-Nagumo model,
\begin{eqnarray}
  u_t &=& au -u^3 - v + \nabla^2 u,  \\
  v_t &=& \varepsilon(u-v)+\delta \nabla^2 v,
\end{eqnarray}
here variables $u$ and $v$ represent the concentrations of the activator and inhibitor, respectively, and $\delta$ denotes the ratio of their diffusion coefficients. The small value $\varepsilon$ characterizes the time scales of the two variables, where $v$ remains approximately constant $v_f$ on the length scale over which $u$ varies. The system described by Eqs.(1) and (2) can be excitable, Turing-Hopf, or bistable type. In this paper the parameter $a$ is chosen such that the system is bistable. The two stationary and uniform stable states are indicated by up state ($u_{+}$,$v_{+}$) and down state ($u_{-}$,$v_{-}$), respectively, and they are symmetric, $(u_{+},v_{+})$=$-(u_{-},v_{-})$. A front (interface) connects the two stable states smoothly. On decreasing $\varepsilon$ the system follows NIB bifurcation that leads to the formation of a couple of Bloch fronts. In the followings, we define a front which jumps from down to up state ($v_{f}$$<$$0$, the planar front velocity $c_{nf}$), and a back which falls from up to down state ($v_{f}$$>$$0$, the planar back velocity $c_{nb}$). Here the front and the back correspond to the two Bloch fronts. So the image of bistable spiral wave is clear: a couple of Bloch fronts (front and back) propagating with opposite velocities enclose a spiral arm, and the front meets the back at the spiral tip at where $v_{f(b)}$$=$$0$ and $u$$=$$v$$=$$0$. In order to differentiate the obtained dense and sparse spiral waves, we define an order parameter $\alpha=|\frac{\lambda_+-\lambda_-}{\lambda_++\lambda_-}|$, here $\lambda_+$ and $\lambda_-$  present the average widths of up state and down state, respectively, which can be served as an indicator of duty ratio of spiral wave in a simple way. In dense spiral case, $\alpha$$=$$0$ which means that the up and down states own identical widths [as Fig. 1 (c) indicated]. In sparse spiral case, if $\lambda_+$$>$$\lambda_-$ we call it Negative Phase Sparse Spiral [NPSS, Fig.1 (d)]. Otherwise if $\lambda_+$$<$$\lambda_-$ we call it Positive Phase Sparse Spiral [PPSS, Fig. 1(e)-(g)].

\indent  Because pattern formations in bistable media are sensitive to the initial conditions and boundary conditions, we adopt two types of fixed initial conditions during numerical simulations as shown in Fig.1 (a) and (b). In order to investigate the transformation process between the dense and sparse spirals, we use Fig.1 (a) [Fig.1 (b)] as the initial condition when increasing (decreasing) $\varepsilon$, which firstly evolves into dense (sparse) spiral. The boundary conditions are taken to be no-flux. The simulation is done in a two-dimensional (2D) Cartesian coordinate system with different grid sizes. A generalized Peaceman-Rachford ADI scheme is used to integrate the above model with a space step $dx$$=$$dy$$=$$0.3$ length unit and a time step $dt$$=$$0.05$ time unit. Unless otherwise noted, our simulations are under the parameter sets: $a$$=$$2.0$, $\delta$$=$$0.1$.
\section{Simulation results and discussion}
\subsection{Bifurcation scenario of bistable spiral waves}
 \begin{figure}[htbp]
  \begin{center}\includegraphics[width=8cm,height=4.7cm]{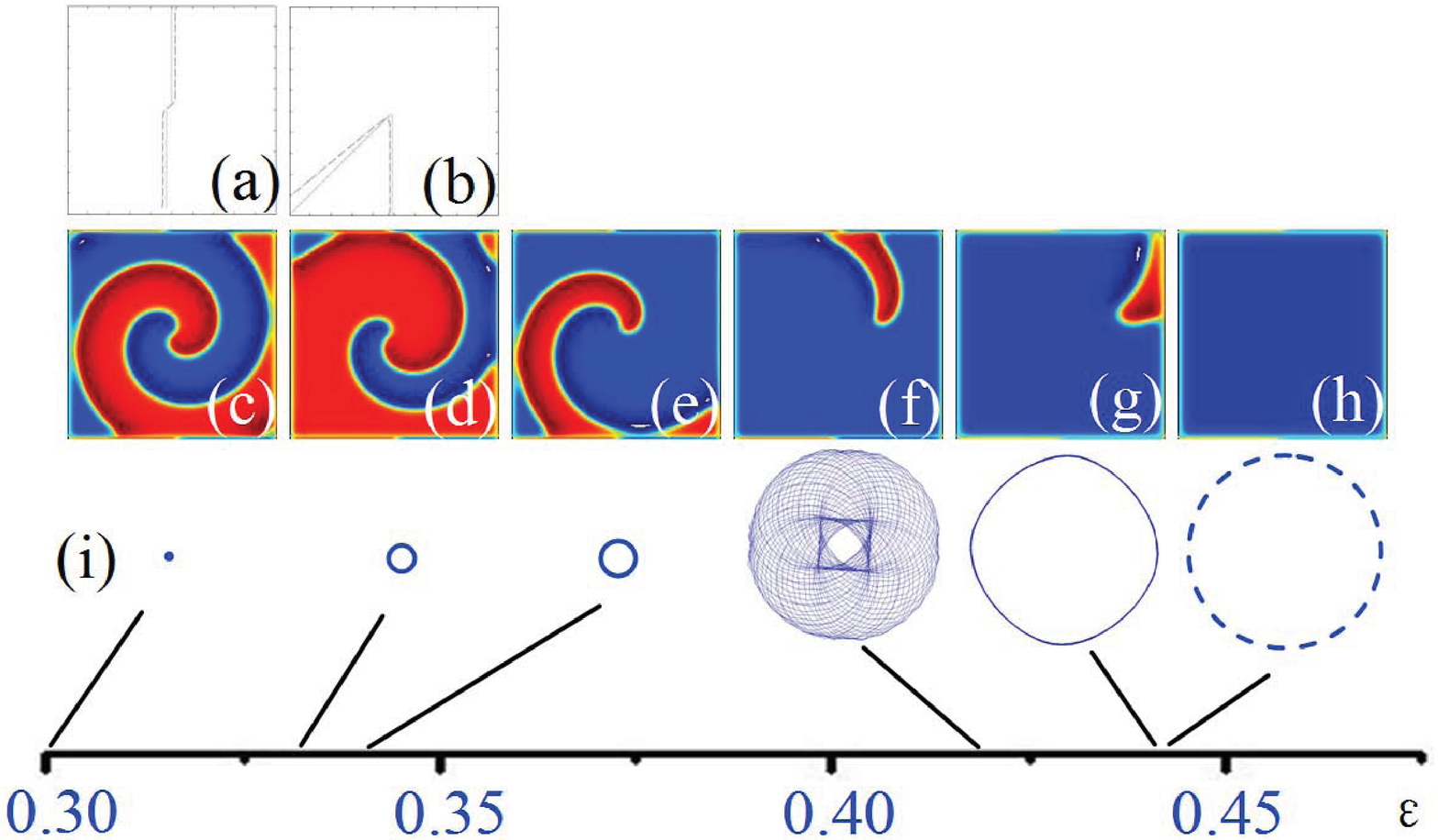}
  \caption{Evolution of spiral waves and tip paths with increasing $\varepsilon$.
 (a) symmetric (b) asymmetric initial conditions. (c) dense spiral, $\varepsilon$$=$$0.3$;
(d) NPSS, $\varepsilon$$=$$0.33$; (e)-(g) PPSS, $\varepsilon$$=$$0.34, 0.42, 0.4385$; (h) uniform state, $\varepsilon$$=$$0.439$; (i) corresponding tip paths. The dash (solid) lines in (a) and (b) present the contour lines $u$=$0$ ($v$=$0$). The dash circle in (i) corresponding to (h) represents a rough tip trajectory of spiral which travels outside the domain. In order to illustrate the meandering of sparse spiral a small domain size is used: $64$$\times$$64$ s.u.}
  \label{1}
  \end{center}
\end{figure}
\indent  Figure 1 shows the bifurcation scenarios of spiral waves and tip trajectories with increasing $\varepsilon$. Here, we use a small domain size $64$$\times$$64$ s.u. in order to illustrate the boundary-induced spiral meandering simultaneously. It can be seen that, upon increasing $\varepsilon$, the system follows dense spiral - sparse spiral - meandering of sparse spiral - sparse spiral - uniform state. When $\varepsilon$$<$$0.33$, the spiral wave is dense [Fig. 1(c)], and the spiral tip is a fixed point. When $\varepsilon$ exceeds the critical value $\varepsilon_{c}$$=$$0.33$ the dense spirals transform into either PPSS or NPSS. The order parameter $\alpha$ increases with $\varepsilon$. After this transformation the tip of sparse spiral begins to travel and traces out a circle with primary radius $r_{1}$ as the spiral wave rotating as indicated in Fig. 1(i). The radius $r_{1}$ increases with $\varepsilon$ as shown in Fig. 2. When $\varepsilon$ approaches roughly $0.404$, we observe the meandering of spiral waves. A secondary circle appears on the trajectory due to the meandering and the primary circle (radius $r_{1}$) orbits the secondary circle (radius $r_{2}$) in anticlockwise direction with frequency $f_{2}$. The primary circle spins about its center in clockwise direction with frequency $f_{1}$. The radius of secondary circle $r_{2}$ reduces upon increasing $\varepsilon$. When the parameter $\varepsilon$ reaches a critical value $0.445$, the meandering of spiral waves disappears and $r_{2}$$=$$0$. If $\varepsilon$ exceeds this value the spiral tip travels outside the domain, so one obtains uniform state (either up state or down state).
\begin{figure}[htbp]
  \begin{center}\includegraphics[width=7cm,height=5cm]{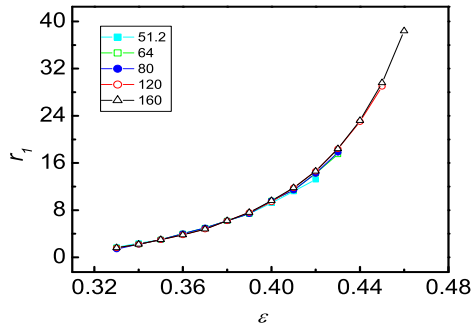}
  \caption{Dependence of the radius $r_{1}$ on parameter $\varepsilon$ under different domain sizes. The numbers in the label present the domain sizes.}
  \label{2}
\end{center}
\end{figure}

\indent  Figure 3 gives a two-parameter numerical bifurcation diagram of spiral waves in a finite domain. The below line $L1$ separates the dense spiral and sparse spiral, and it keeps constant while changing the domain sizes. There exists a parameter range corresponding the meandering of sparse spiral as indicated by the gray region in Fig. 3. This region changes with domain sizes. If the system is large enough, for example $160$$\times$$160$ s.u., this region reduces to a line which separates the sparse spiral and the uniform state. So the observed meandering of sparse spiral is induced by the boundary effect.

\begin{figure}[htbp]
  \begin{center}\includegraphics[width=7cm,height=5cm]{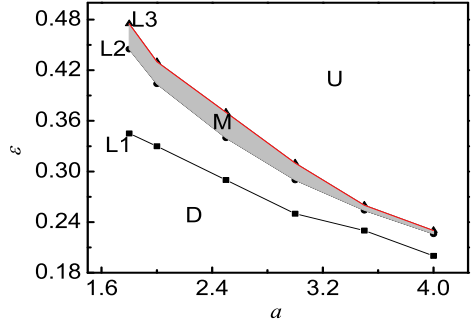}
  \caption{Two-parameter numerical bifurcation diagram of spiral waves. The region between lines $L1$ and $L3$ presents sparse spirals. The gray region M indicates meandering sparse spirals. The regions $D$ and $U$ denote Dense spirals and Uniform states, respectively. Domain size: $64$$\times$$64$ s.u.}
  \label{3}
\end{center}
\end{figure}

\begin{figure}[htbp]
  \begin{center}\includegraphics[width=7cm,height=5cm]{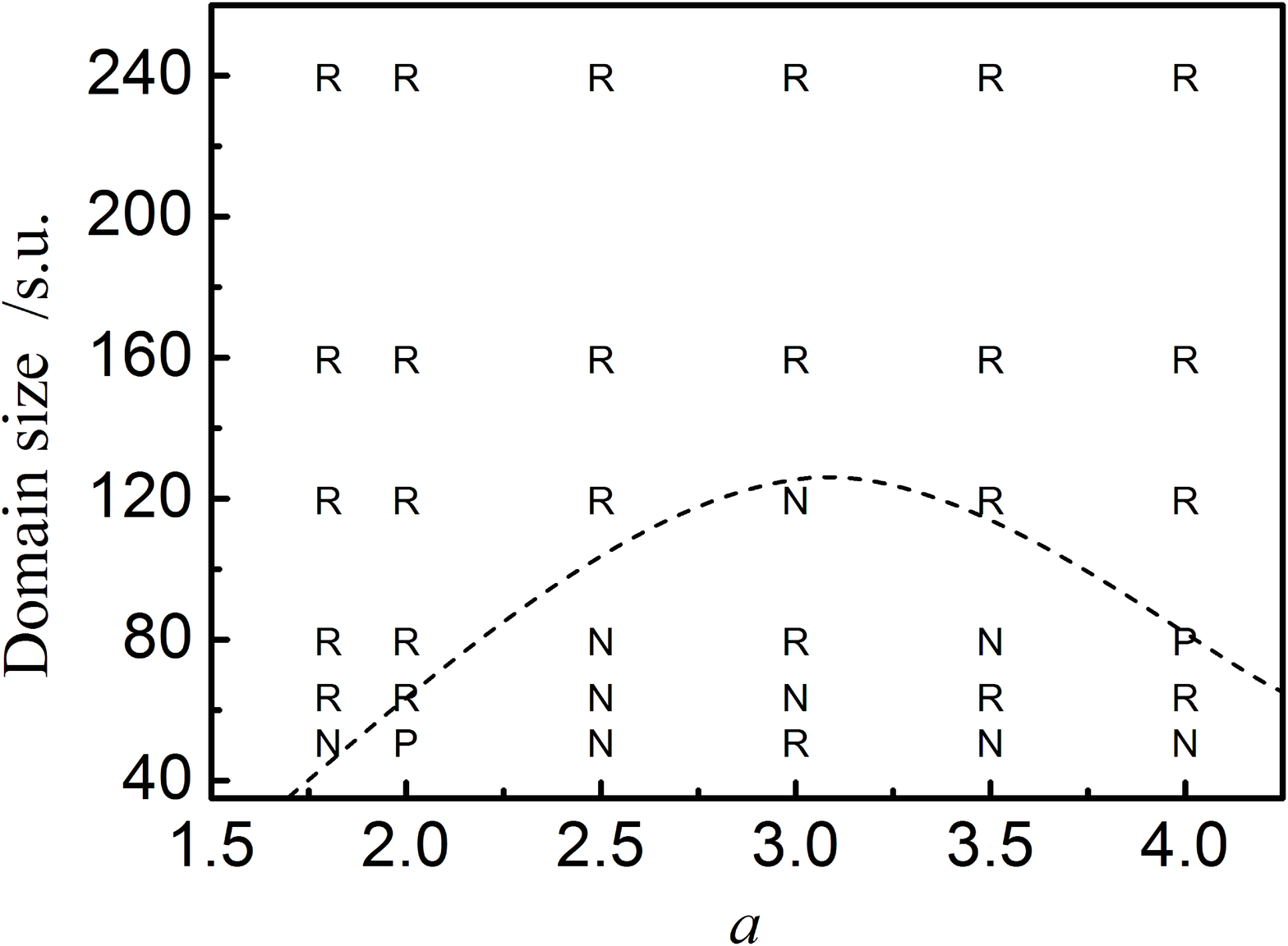}
  \caption{Phase diagram of the sparse spirals spanned by the domain size and the parameter $a$. This diagram presents the choice of NPSS or PPSS after the transformation as increasing $\varepsilon$. R denotes Ruleless choice. P (N) indicates PPSS (NPSS). The dashed line divides roughly the states of sparse spirals after the transformation into two regions, above which the choice is always Ruleless.}
  \label{4}
  \end{center}
\end{figure}

\indent As indicated in Fig. 1, the dense spiral can transit into either PPSS or NPSS as increasing $\varepsilon$. In order to study the choice of NPSS or PPSS after the transformation, we do extensive simulation with different parameter sets: $\varepsilon$, $a$, and domain sizes. For example, given fixed parameters: domain size $64$$\times$$64$ s.u., $a$$=$$2$ as shown in Fig. 1, we try to connect the spiral type with the parameter $\varepsilon$. However, at different $\varepsilon$, the dense spiral transforms into different types of sparse spirals. We don't find rule about how the parameter $\varepsilon$ determines the type of sparse spiral. Here we denote this transformation as Ruleless choice. Nevertheless, at some specific parameters, the dense spiral always transits into one type of sparse spiral, either PPSS or NPSS. For example, given domain size $64$$\times$$64$, $a$$=$$3$, the dense spiral always transits into NPSS at different $\varepsilon$. We give a phase diagram of the spiral types as shown in Fig. 4, in which R represents Ruleless choice, P (N) indicates the PPSS (NPSS). It is shown that if the domain size is large enough (roughly, the region above the dash line) dense spiral evolves into either PPSS or NPSS as changing $\varepsilon$. The transformed states PPSS and NPSS can be regarded as two stable states of system after the bifurcation. Without forcing (for large domain sizes), the choice is ruleless due to the spontaneous symmetry breakdown. For smaller domain size, however, this choice tends to be unique after the transformation. It is obvious that this choice originates from boundary effects.

\indent  We want to mention that Fig. 4 shows a phase diagram with different parameters. However, given a set of parameters, the choice is deterministic. For example, given domain size $64$$\times$$64$ s.u., $a$$=$$2.0$, $\varepsilon$$=$$0.33$, the initial dense spiral will transform into NPSS as shown in Fig. 1 (d). So, if decreasing $\varepsilon$ from $0.33$ to $0.3$ then increasing to $0.34$, one can turn a NPSS into dense spiral then into PPSS. This process changes the duty ratio of spiral waves. It is helpful to adjusting the duration of a cardiac action potential to prevent the atrial fibrillation. \cite{Nattel}

\indent Sparse spiral can also evolve into dense spiral as decreasing $\varepsilon$. From the given asymmetric initial condition [Fig. 1(b)], it first develops into sparse spiral. When $\varepsilon$$<$$0.315$ this sparse spiral further evolves into dense spiral. We want to mention that, as shown in Fig. 1 when increasing $\varepsilon$, the transition from dense to sparse spiral occurs at $\varepsilon$$=$$0.33$. So, this transformation is bistable with respect to $\varepsilon$, and originates from a subcritical bifurcation. This transformation process is very different from that in excitable media, \cite{Krinsky, Kessler} in which the transformation between sparse and dense spirals is smooth. The present phenomena attribute to the intrinsic characteristics of the bistable system.

\subsection {Transformation details from dense spiral to sparse spiral}

\indent Because this transformation is bistable, here, we focus on the transformation details from dense spiral to sparse spiral. When the parameter $\varepsilon$ has exceeded $\varepsilon_{c}$, the system with the given symmetric initial condition [Fig. 1 (a)] evolves first into dense spiral, then into sparse spiral. We study the transformation in detail by measuring the intensity change of variable $u$ and $v$ at a fixed point far away from the spiral tip and the domain boundary as shown in Fig. 5. It can be seen that the period of sparse spiral $T_{1}$ is always larger than that of dense spiral $T_{0}$. There exists a critical period $T_{c}$ (interval of $u_{+}$) at which the transformation from dense spiral to sparse spiral occurs. It is found that the differences between $T_{0}$ and $T_{c}$ are related to the choice of PPSS or NPSS,
\begin{displaymath}
\left\{ \begin{array}{ll}
T_0 < T_c, & \textrm{PPSS},\\
T_0 > T_c, & \textrm{NPSS}.
\end{array} \right.
\end{displaymath}
\begin{figure}[htbp]
  \begin{center}\includegraphics[width=7cm,height=5cm]{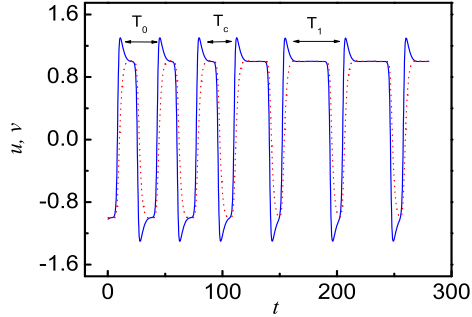}
  \caption{Time sequence of variable $u$ and $v$ around the transformation point for NPSS.  Solid line and dash line present $u$ and $v$, respectively. $T_{0}$ and $T_{1}$ are the periods of dense spiral
and sparse spiral, respectively. $T_{c}$ indicates a critical period.}
  \label{5}
\end{center}
\end{figure}

\indent The appearance of $T_{c}$ indicates the change in velocities of both the front and the back. For the case of small $\varepsilon$ and $\delta$, Eqs. (1) and (2) reduce to

\begin{eqnarray}
  u_t &=& au - u^3 - v_i + \nabla^2 u, \\
  v &=& v_i  (i = f, b),
\end{eqnarray}
here $v_{i}$ is the value of the inhibitor at the front or the back. From Eqs.(3) and (4) we can obtain the velocity of the planar front (back) $c_{ni}$=$-\frac{3}{\sqrt{2}a}$$v_{i}$($i$=$f$,$b$). So the value $v_{i}$ determines uniquely the velocity of front (back). For
dense spiral, the front and the back are symmetry except an angle separation of $\pi$ and travel at the same speed, $|v_{b}|$$=$$|v_{f}|$. The interaction between the front and the back is negligible. Undergoing a subcritical bifurcation the symmetry between the front and the back breaks down. Here, if $|v_{f}|$$>$$|v_{b}|$, the front speeds up, so the measured critical period $T_{c}$ would be less than $T_{0}$. The accelerated wave front approaches and interacts with the back to form a NPSS. On the contrary, if $|v_{b}|$$>$$|v_{f}|$ the back speeds up, so $T_{c}$$>$$T_{0}$ and a PPSS will appear.

\indent  The velocities of both the front and the back are also restricted by the eikonal relation which amends the velocity difference between the front and the back induced by the symmetry breakdown,
\begin{equation}\label{7}
    c_i = c_{ni}(v_i) - D\kappa_i.
\end{equation}
here, $\kappa_i$ indicates the local curvature. For NPSS (PPSS) the front and the back are asymmetric and the local curvature $\kappa_f$ of the front is larger (smaller) than that $\kappa_b$ of the back far away the tip. So, the final velocities of the front and the back are consistent, which leads to stable spiral wave.

\indent In the simulation we find that the period difference between the dense and sparse spirals and the radius $r_{1}$ of primary circle satisfy the following approximate relation,
\begin{equation}\label{4}
    T_1 - T_0 =8 r_1.
\end{equation}
\indent  If $r_{1}$$=$$0$, $T_1$ reduces to $T_0$, and it corresponds to dense spiral. Because the velocities of both the front and the back decrease with $\varepsilon$, $T_{0}$ and $T_{1}$ increase with $\varepsilon$. The closer the parameter $\varepsilon$ is to $\varepsilon_{c}$, the longer it needs to evolve from the initial condition to sparse spiral. This attributes to the critical slowing down near the bifurcation point $\varepsilon_{c}$.

\indent  In the present case, the system is symmetric and the symmetry breakdown of the front and the back originates from a subcritical bifurcation, which differs from that in Ref. 21. They obtained the dense and sparse spirals in symmetric and asymmetric bistable systems, respectively, in the limit of $\varepsilon$/$\delta$$\ll$$1$. In that case, the symmetry breakdown originates from a saddle-node bifurcation induced by a constant in the model.
\begin{figure}[htbp]
  \begin{center}\includegraphics[width=7cm,height=8cm]{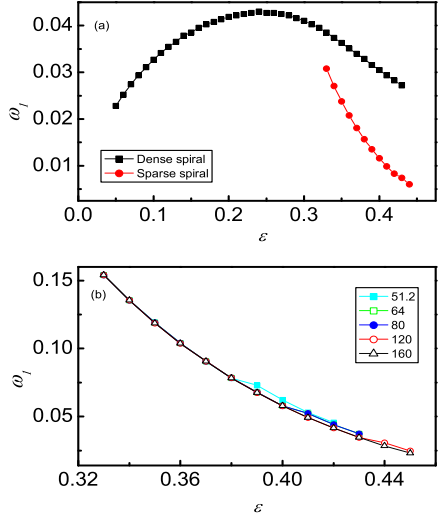}
  \caption{Dependence of rotation frequency of spirals on the parameter $\varepsilon$. In (a) the filled square (cycle) denotes dense (sparse) spiral. Domain size: $64$$\times$$64$ s.u. (b) shows the rotation frequencies of sparse spirals under different domain sizes. The curves for small domain sizes diverge when $\varepsilon$ is larger than about $0.38$.}
  \label{6}
\end{center}
\end{figure}

\indent Figure 6 (a) shows the dependence of rotation frequency of dense and sparse spirals on the parameter $\varepsilon$. The filled square (cycle) denotes dense (sparse) spiral. It can be seen that the rotation frequency of dense (sparse) spiral varies nonlinearly, nonmonotonously (monotonously) with respect to $\varepsilon$. The frequency of sparse spiral disappears for $\varepsilon$$<$$0.33$, and is always smaller than that of dense spiral, which is due to the rigid rotation.

\subsection{Meandering of sparse spiral induced by boundary effects}

\indent  Figure 6 (b) shows the dependence of average primary frequency $\omega_{1}$ on the parameter $\varepsilon$ under different domain sizes. It can be seen that for small $\varepsilon$ these curves are superposition. But for large $\varepsilon$ they begin to diverge. The onset of diverge increases with domain sizes and indicates the startpoint of meandering spiral. The rotation frequency increases with decreasing domain sizes at larger $\varepsilon$. This phenomenon agrees well with the experimental observation in Ref. 28. The radius $r_{1}$ of primary circle decreases with domain sizes for larger $\varepsilon$ as shown in Fig. 2, which attribute to the interaction between the spiral tip and the boundary. In addition, near the onset, the sparse spiral begins to meander and the radius $r_{2}$ of the secondary cycle tends to be infinite. But due to the boundary effects, $r_{2}$ is limited to a distance away from the boundary about $10$ s.u.. With increasing $\varepsilon$, $r_{2}$ decreases dramatically and then tends to be zero at certain bifurcation point. Near this point the radius $r_{2}$ scales approximately as the square-root of the distance from this bifurcation point. This means that the boundary-induced meandering originates from a supercritical bifurcation.

\indent  From the simulation for different domain sizes and $\varepsilon$ we can give an experiential condition for meandering of sparse spirals,
\begin{equation}\label{9}
    L/\lambda\in(0.5, 0.7),
\end{equation}
where $L$ is the domain size and $\lambda$ is the wavelength of sparse spiral. It demonstrates that meandering of spiral occurs when the wavelength of spiral reaches at least twice of the domain size.

\indent We only observed the meandering of sparse spirals with outward petals even by changing the chiralities of spirals. No meandering spiral with inward petal is found in simulations. The observed characteristics of meandering spirals differ greatly from that observed in excitable media in Refs. 29, 30 in which hypocycloidlike orbit transits into epibycloidlike orbit as changing a control parameter continuously. That meandering with both outward and inward petals originates from a pair of secondary Hopf bifurcations. But in the present work, from the description of Fig. 3, it can be seen that the obtained meandering of sparse spiral originates from boundary effect rather than regular meandering instability. By investigating the linear stability analysis of spirals, B\"{a}r and co-workers \cite{Bar3} observed meandering-Hopf bifurcation (regular meandering which contains a couple of Hopf bifurcations as Barkley reported) and boundary-Hopf bifurcation (boundary-induced Hopf bifurcation which contains only one Hopf bifurcation point). The boundary-Hopf bifurcation has two characteristics: 1) it occurs only when the domain size is smaller than a critical value; 2) it contains only one bifurcation point. Our results, as shown in Fig. 1-3, satisfy the two conditions. So, the meandering spirals in our case result from boundary-induced Hopf bifurcation. If there exists regular meandering for bistable spiral, it is impossible to make a simulation because it tends to be uniform state when $\varepsilon$ is large enough.

\section{CONCLUSION AND REMARKS}
\indent We have studied the transformation between dense and sparse spirals based on a bistable FitzHugh-Nagumo model and investigated a novel meandering of sparse spiral with only outward petals. The dense spiral and the sparse spiral can transform each other via a subcritical bifurcation which differs from that in excitable media. \cite{Krinsky, Kessler} The dense spiral can transit into either PPSS or NPSS. We can find a way to turn a PPSS into dense spiral then into NPSS by decreasing $\varepsilon$ and then increasing it again, which provides a simple method to adjusting the duration of a cardiac action potential to prevent the atrial fibrillation. \cite{Nattel} By using different domain sizes we have studied the meandering of sparse spiral and given an experiential condition for spiral meandering induced by the boundary effects $L$/$\lambda$$\in$($0.5$, $0.7$). The observed meandering of sparse spiral with only outward petal originates from a Hopf bifurcation induced by the boundary effects.

\indent Although the present FitzHugh-Nagumo model isn't a realistic model describing chemical reaction, it has successfully explained many pattern formations in ferrocyanide-iodate-sulfite reactions,\cite{Li2, Hagberg0, Hagberg1, Szalai, Lee1, Lee2, Lee3} such as the spiral, the self-replicating spots, and the labyrinthine pattern which occur on the left, near, and on the right of the NIB bifurcation point, respectively. In the present work we studied the transformation between dense and sparse spirals on the left of this bifurcation point. We hope that our results can be observed in a ferrocyanide-iodate-sulfite reaction.

\section{ACKNOWLEDGMENTS}
\indent  This work is supported in part by Hong Kong Baptist University and the Hong Kong Research Grants Council. Y. F. He also acknowledges the National Natural Science Foundation of China with Grant No. 10775037, 10947166, and the Research Foundation of Education Bureau of Hebei Province, China (Grant No. 2009108).

\end{document}